\documentclass{sig-alternate-2013}

\newfont{\mycrnotice}{ptmr8t at 7pt}
\newfont{\myconfname}{ptmri8t at 7pt}
\let\crnotice\mycrnotice%
\let\confname\myconfname%

\permission{Permission to make digital or hard copies of all or part of this work for personal or classroom use is granted without fee provided that copies are not made or distributed for profit or commercial advantage and that copies bear this notice and the full citation on the first page. Copyrights for components of this work owned by others than ACM must be honored. Abstracting with credit is permitted. To copy otherwise, or republish, to post on servers or to redistribute to lists, requires prior specific permission and/or a fee. Request permissions from Permissions@acm.org.}
\usepackage{verbatim}
\usepackage{hyperref}
\usepackage{amssymb}
\usepackage{amsmath}
\usepackage{amsfonts}
\usepackage{graphicx}
\usepackage{enumitem}
\usepackage{color}
\usepackage{algorithm}
\usepackage{algorithmicx}
\usepackage{float}
\usepackage{algpseudocode}
\usepackage{multirow}
\usepackage{epstopdf}
\usepackage{graphicx}
\usepackage{subfigure}

\usepackage{listings}
\usepackage{color}

\definecolor{dkgreen}{rgb}{0,0.6,0}
\definecolor{gray}{rgb}{0.5,0.5,0.5}
\definecolor{mauve}{rgb}{0.58,0,0.82}

\hypersetup{
    colorlinks,
    citecolor=blue,
    filecolor=black,
    linkcolor=blue,
    urlcolor=blue,
    pdfauthor={},
    pdfsubject={Blog Recommendation in Tumblr},
    pdftitle={Blog Recommendation in Tumblr},
    pdfkeywords={Collaborative Filtering, Content-based Recommender System, Cross-domain Recommendation}
}
\newdef{definition}{Definition}
\setlist[itemize]{noitemsep, topsep=0pt}
\newfloat{algorithm}{t}{lop}

\begin{document}

\title{The Apps You Use Bring The Blogs to Follow}

\numberofauthors{3}
\author{
\alignauthor
Yue Shi\\
       \affaddr{Yahoo! Research}\\
       \affaddr{Sunnyvale, CA, USA}\\
       \email{yueshi@acm.org}
\alignauthor
Erheng Zhong\titlenote{Work done when the author was at Yahoo!.}\\
       \affaddr{Baidu Big Data Lab}\\
       \affaddr{Sunnyvale, CA, USA}\\
       \email{zhongerheng@baidu.com}
\alignauthor Suju Rajan\titlenote{Work done when the author was at Yahoo!.}\\
       \affaddr{Criteo Labs}\\
       \affaddr{Palo Alto, CA, USA}\\
       \email{s.rajan@criteo.com}
\and
\alignauthor Liang Dong\\
       \affaddr{Yahoo! Tumblr}\\
       \affaddr{New York, NY, USA}\\
       \email{ldong@tumblr.com}
\alignauthor Hao-wei Tseng\\
       \affaddr{Yahoo! Tumblr}\\
       \affaddr{New York, NY, USA}\\
       \email{haowei@tumblr.com}
\alignauthor Beitao Li\\
       \affaddr{Yahoo! Tumblr}\\
       \affaddr{New York, NY, USA}\\
       \email{beitao@tumblr.com}
}

\maketitle
\begin{abstract}
We tackle the blog recommendation problem in Tumblr for mobile users in this paper. Blog recommendation is challenging since most mobile users would suffer from the cold start when there are only a limited number of blogs followed by the user. Specifically to address this problem in the mobile domain, we take into account mobile apps, which typically provide rich information from the users. Based on the assumption that the user interests can be reflected from their app usage patterns, we propose to exploit the app usage data for improving blog recommendation. Building on the state-of-the-art recommendation framework, Factorization Machines (FM), we implement app-based FM that integrates app usage data with the user-blog follow relations.  In this approach the blog recommendation is generated not only based on the blogs that the user followed before, but also the apps that the user has often used. We demonstrate in a series of experiments that app-based FM can outperform other alternative approaches to a significant extent. Our experimental results also show that exploiting app usage information is particularly effective for improving blog recommendation quality for cold start users.

\end{abstract}

\noindent
{\bf Categories and Subject Descriptors:} H.4 {[Information Systems Applications]}: {Miscellaneous}

\noindent
{\bf General Terms:} Design, Experimentation


\begin{figure}[t]
\begin{small}
\centering {
{{\includegraphics[width=5cm]{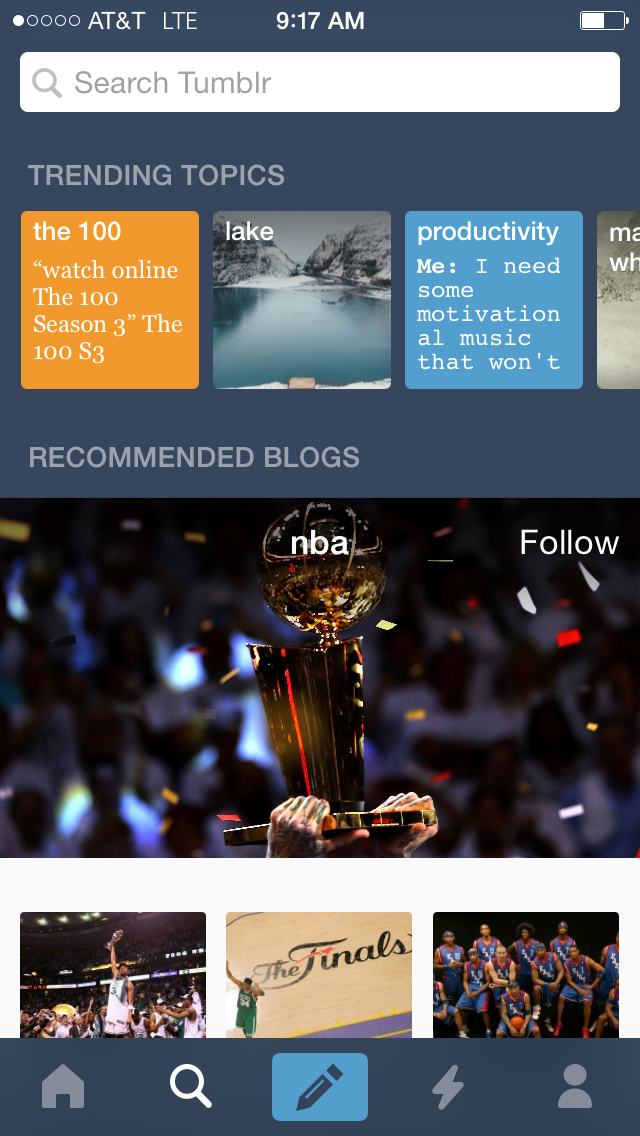}}}
}\caption{An example of Blog recommendation in Tumbr from a mobile device.} \label{fig:screenshot}
\end{small}
\end{figure}

\section{Introduction}\label{sec:intro}
Tumblr\footnote{https://www.tumblr.com/} is a major social media site for microblogging. Similar to Twitter\footnote{https://twitter.com}, the social network structure in Tumblr is directed, i.e., a user can follow a blog without the followee's confirmation, nor being followed back. However, Tumblr is known to be different from Twitter in the sense that it is much richer in text and images~\cite{Chang2014WIT}. One key aspect of the growth of Tumblr is to engage the users with more relevant blogs to follow. Thus, one of the most critical technical challenges in Tumblr becomes how to make blog recommendations that the users are likely to follow. As mobile devices become ubiquitous, this challenge calls for special attention, since it has a great impact on the mobile experience of Tumblr users. For example, for a user who mainly accesses social media sites from mobile, it would be very difficult to make blog recommendations to this user if she is new to Tumblr. Failing to find interesting blogs quickly would cause user abandoning the site in an early stage, and thus, being harmful for the growth of Tumblr in mobile. Figure~\ref{fig:screenshot} shows an example of the blog recommendation module in Tumblr from a mobile device. A user can scroll the screen to see blogs recommended in the "Recommended Blogs" section. If the user is interested in one blog, she can click the "Follow" tap to follow the blog and then she will be notified all the updates in the blog. Undoubtedly, having user follow blogs is the key metric that drives our evaluation on the performance of blog recommendation. 

Our goal in this paper is to tackle the blog recommendation problem in the mobile domain.
We shall emphasize two critical issues that we face in order to successfully address this problem.
\begin{itemize}
	\item \textbf{Scalability.} Tumblr is one of the largest social networks in the world. Blog recommendation in Tumblr involves hundreds of millions users and tens of billions interactions between users and blogs. Therefore, we need a recommendation system that is able to process the data at such a scale. In addition, we may exploit even richer source of information that would make the scale of the data much larger. For this reason, it is expected that our recommendation framework is highly scalable to very large datasets.
	\item \textbf{Cold Start.} One naive way to make blog recommendation is to adopt the conventional collaborative filtering (CF) methods, i.e., to derive recommendations based on the user-blog matrix. In other words, we can make use CF methods to generate blog recommendations for a user based on what the user has followed before. The performance of such methods can be very limited for the users who only have a small number of followed blogs, so called the cold start problem. Since the long tail is a deterministic property of social networks, we need to make specific contributions to improve the recommendation performance for the tail users. One straightforward consideration is to exploit the demographic information of the mobile users, in addition to the limited following history. Note that, we suppose, in most cases, the basic demographic information of users is available from mobile devices. However, the basic demographic information may not be discriminative enough for serving personalized blog recommendation to individual users. For example, the gender of the user may only be able to distinguish a limited categories of blogs for the user, while not being able to capture a wide scope of the user's interest. For this reason, it is desired that we exploit extra rich sources of information that can help to infer the underlying user interests from mobile devices.
\end{itemize}

In this paper, we propose a blog recommendation system for Tumblr mobile that specifically addresses the aforementioned issues. We exploit distributed implementations of latent factor models, in particular, Factorization Machine (FM)~\cite{Rendle2012FML}, which can scale to the large datasets as in Tumblr. In addition, the intrinsic property of FM allows us to incorporate rich sources of information beyond the user-blog interactions. To address the cold start problem, we propose to use the app usage information of mobile users to infer the user interest. As the experience of mobile users is dominated by the apps that they use, it is reasonable to learn user interest from mobile apps.
Our assumption is that the apps that a user often uses may indicate the topic interests of the user, which can be generalized to the domain of blog recommendation. We collected a large set of app usage data of mobile users, and demonstrate the correlation between app usage and blog topics. We further conduct a series of experiments showing that great improvement can be achieved for blog recommendation in Tumblr by exploiting app usage information. 

Our contributions in this work can be summarized in twofold:
\begin{itemize}
 \item We analyze and identify in the mobile domain the relationship between Tumblr blogs and mobile apps, brining new insights to the application of blog recommendation.
 \item We propose an approach by using FM to exploit app usage data for blog recommendation and show its effectiveness through experiments on a large scale dataset.
\end{itemize}

The paper is organized as follows. In Section~\ref{sec:rw}, we review related work, and position our work with respect to it. In Section~\ref{sec:appblogdata}, we analyze the data in the problem domain and discuss our motivation. After that, in Section~\ref{sec:model} we present the details of the recommendation approaches for blog recommendation, especially for exploiting apps in the FM framework. Then, in Section~\ref{sec:experiment}, we demonstrate the results of our experiments, followed by Section~\ref{sec:conclusion} that concludes the paper.

\begin{figure*}[t]
\begin{small}
\centering \mbox{
\subfigure[Distribution of blogs]{\scalebox{1.0}{\includegraphics[width=\columnwidth]{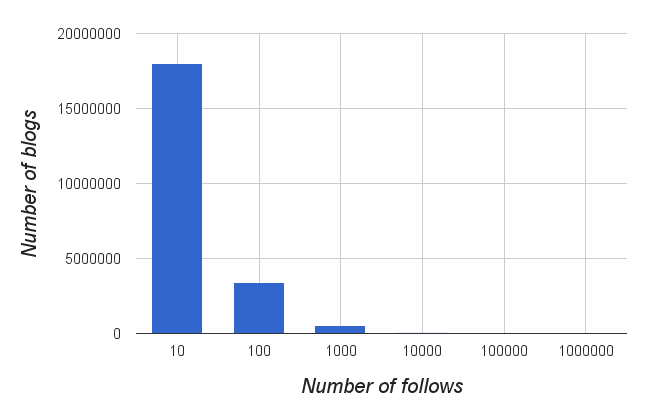}}\label{fig:distr:blog}}
\subfigure[Distribution of blog users]{\scalebox{1.0}{\includegraphics[width=\columnwidth]{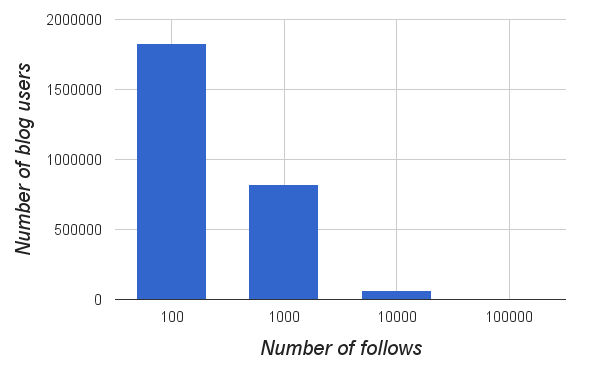}}\label{fig:distr:bloguser}}
}\caption{Distribution of blog following data} \label{fig:blogdistr}
\end{small}
\end{figure*}
\begin{figure*}[t]
\begin{small}
\centering \mbox{
\subfigure[Distribution of apps]{\scalebox{1.0}{\includegraphics[width=\columnwidth]{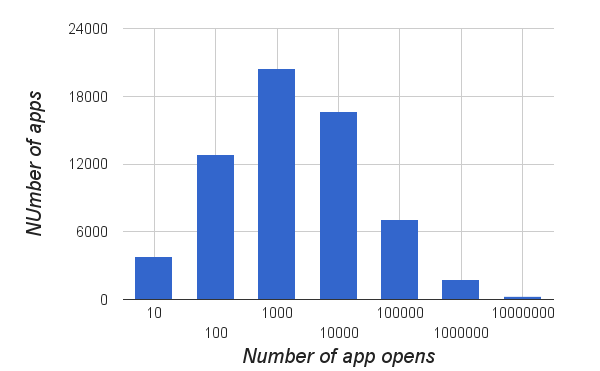}}\label{fig:distr:app}}
\subfigure[Distribution of app users]{\scalebox{1.0}{\includegraphics[width=\columnwidth]{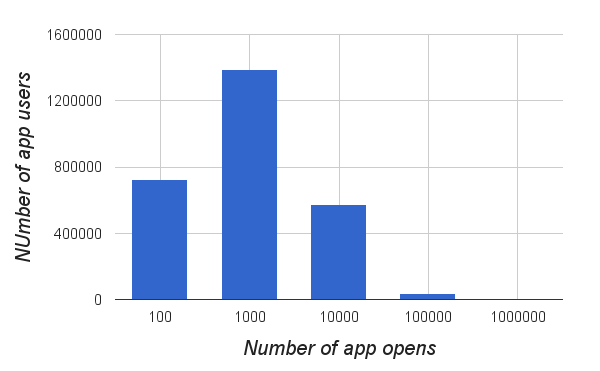}}\label{fig:distr:appuser}}
}\caption{Distribution of app usage data} \label{fig:appdistr}
\end{small}
\end{figure*}

\section{Related Work}\label{sec:rw}
Our work is related to several aspects in the area of recommender systems. In this section, we discuss three key research topics that are closest to our work in this paper.

\subsection{Social Recommendation}
Blog recommendation in Tumblr falls into the intersection between recommender systems and social networks, a particular area known as \textit{social recommendation}~\cite{Guy2014SRS,yang2014ASO,Shi2014CFB}. Great efforts have been devoted to social recommendation in the research community, in which conventional collaborative filtering approaches, such as matrix factorization and random walks, are extended to incorporate the information from social networks for further improving recommendation performance~\cite{Konstas2009SNC,Ma2011RSS}. As an industry practice, our work is close to the work reported from other social network companies, such as LinkedIn feed recommendation~\cite{Agarwal2014ARL,Agarwal2015PLF} and follower recommendation at Twitter~\cite{Gupta2013WFS}. The difference lies in that 1) we target a new problem domain, i.e., blog recommendation in Tumblr; and 2) we focus on leveraging external auxiliary information to address the cold start problem.

\subsection{Latent Factor Models}
The recommendation framework we build on in this work is \textit{Factorization Machines} (FM)~\cite{Rendle2012FML}, which is a particular class of latent factor models. Latent factor models have shown their effectiveness for improving recommendation performance through several public competitions in the past years, such as in Netflix Prize competition~\cite{Koren2009MFT} and KDD CUP~\cite{Chen2012SVD}. The core concept of latent factor models is to learn low rank representations of users and items so that they are indicative for users' preference to items.
FM was proposed as a generalized latent factor model, one of its key advantages being able to incorporate additional information sources in a uniform fashion~\cite{Rendle2012FML,Rendle2013SFM}. Since its inception, FM has been widely recognized and adopted in a number of use cases, such as context-aware recommendation~\cite{Rendle2011SCR} and cross-domain recommendation~\cite{Loni2014CDC}. Our blog recommendation system is built on FM that allows us to exploit auxiliary information sources in addition to the user-blog interactions. However, we implemented FM in a distributed fashion based on our internal infrastructure so that it can scale to very large datasets. Moreover, our focus is application-specific for blog recommendation, while not on improving the FM modeling framework. 

\subsection{Cross-Domain Recommendation}
Our work is also related to the research topic of cross-domain recommendation, which generally refers to the recommendation scenarios that involve multiple application domains~\cite{Shi2014CFB}. Prior work has shown that sharing knowledge between different product domains can bring up mutual benefits. For example, the users may have similar interest in movies and books~\cite{Li2009CMA}. It is also shown that auxiliary information sources can be exploited as intermediate for linking different domains, e.g., tags may be common between the book domain and the movie domain~\cite{Shi2011TAB}, and users may have similar rating patterns across different product domains~\cite{Li2009CMA,Loni2014CDC}. We consider that external information sources are critical for addressing the cold start problem for blog recommendation. In particular, our work is specialized in the mobile domain, in which we exploit the rich user feedback from their interactions with mobile apps. Our assumption is that the apps that a user often uses can indicate her underlying interests, which can be leveraged for blog recommendation in Tumblr.

\section{Apps and Blogs}\label{sec:appblogdata}
In this section, we analyze the characteristics of blog following data and app usage data, which motivate our proposed solution of exploiting the users' app usage behavior to improve blog recommendation in Tumblr. We collected a sampled dataset of blog following in Tumblr, containing a bipartite graph that shows which user follows which blog. The detail of the dataset is presented in Section~\ref{sec:experiment}, where we demonstrate our experimental evaluation. The app usage dataset is collected internally from the user action logs in mobile devices for one week. For the concern of business confidentiality, we can not disclose the detail of the process of collecting the app usage data in this paper. Since our target is on blog recommendation, we neglect the users who have no blog following information. In addition, for each user we only take into account up-to 10 apps that she most frequently uses. We leave more detail of the dataset in Section~\ref{sec:experiment}.

Figure~\ref{fig:distr:blog} shows the blog distribution in our dataset. As expected, the distribution follows a power law, which indicates that a small number of blogs are followed by a great number of users, while there are massive blogs that are followed by very few users. For example, in this dataset we observe that there are more than 10 million blogs followed by less than 10 users, while there are less than 10 blogs followed by more than 1 million users. This characteristic implies that it would be very challenging to recommend the tail blogs, due to the high data sparseness. In addition, we also observe a power law distribution for the users, as in Figure~\ref{fig:distr:bloguser}. Most users would only have been following a limited number of blogs, resulting in a natural cold start situation for recommending personalized blogs to individual users.

On the other side, we show in Figure~\ref{fig:distr:app} that the distribution of the apps used by the users is close to a normal shape. It indicates that the majority of apps are shared across the average of the user population, which means that apps might play a better role of carrying over user interest than blogs. Moreover, in Figure~\ref{fig:distr:appuser}, we can see that the situation of cold start users from the app's point of view is substantially alleviated, compared to the blog's point of view as in Figure~\ref{fig:distr:bloguser}. As such, it is a reasonable strategy to exploit apps that the users use to improve blog recommendation in Tumblr.

\begin{figure}[t]
\begin{small}
\centering {
{{\includegraphics[width=8cm]{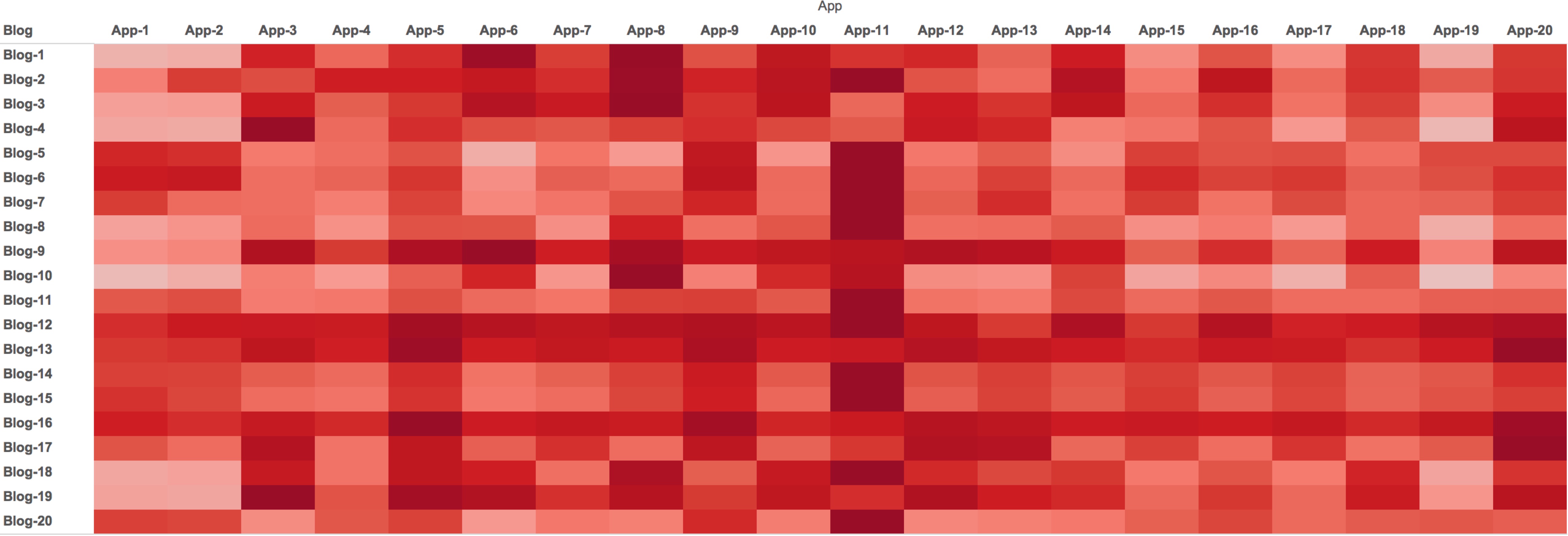}}}
}\caption{Similarity between popular apps and popular blogs.} \label{fig:appblog}
\end{small}
\end{figure}

In order to further understand the relationship between the apps that a user has and the blogs that the user follows, we investigate the correlation between apps and blogs, by means of measuring the cosine similarity between the two based on their users. In other words, if an app is considered similar to a blog, it means that probably a lot of users who use that app also follow that blog. For the convenience of visualization, we focus on the top 20 most popular apps and the top 20 most popular blogs (popular in the sense of the number of users that uses or follows), and show their similarities in Figure~\ref{fig:appblog}. As can be seen, the apps do share similarity with the blogs, indicating that it is reasonable to exploit apps for connecting users to blogs. In addition, we can also see that even among the top popular apps and blogs, they do not have consistent similarity pattern. In other words, some apps may be more similar to one blog, while some other apps may be more similar to another. Therefore, it is also reasonable to believe that mobile apps are sufficiently discriminative for indicating user interests. In the following section we will present the technical detail of the recommendation approaches, in which we particularly focus on exploiting app usage data.

\section{Blog Recommendation Model}\label{sec:model}

As shown above, there are strong correlations between users' app usage logs and their blog following patterns. In this section, we present in this section two algorithms, one baseline approach known as Item-based Collaborative Filtering (ItemCF)~\cite{Sarwar2001ICF}, and one proposed approach built on Factorization Machines (FM)~\cite{Rendle2009} which demonstrates how to utilize app usage logs for improvements of blog recommendations.

\subsection{Problem Formulation}
Let $U$ denote the user set, $V$ denote the blog set, $m$ be the number of users and $n$ be the number of blogs. The following graph can be defined as a sparse matrix $R\in \mathbf{R}^{m*n}$, where $R_{ij} \in \{1, ?\}$, $R_{ij}=1$ denotes that user $u_i\in U$ followed blog $v_j\in V$ and $R_{ij}=?$ means unobserved. In addition, we have side information of users, i.e., their app usage logs $A\in \mathbf{R}^{m*c}$, where $c$ is the number of apps. Typically, $A$ is a sparse matrix as well since users only use a few of apps. Generally, the task of blog recommendation is to predict which users will follow which blogs based on the observed following graph and available side information. Formally, we aim to build models to approximate the observed data by minimizing some loss functions with regularizations.
\begin{equation}
	\min_{F} \Omega(R, A, D|F) + \lambda\mathcal{R}(F)
\end{equation}
where $\lambda$ is the regularization parameter. As follows, we introduce two common algorithms to construct the model $F$.

\subsection{Item-based CF}
As introduced in~\cite{Sarwar2001ICF}, ItemCF (also known as KNN) has been widely used in various recommendation systems. Its basic idea is to construct the similarities among different items, and then based on users interacted items, other similar items can be recommended to users. Formally, let $\mathbf{u}_i\in \mathbf{R}^n$ denote the users' interactions with items, and $S\in \mathbf{R}^{n*n}$ denote the similarity matrix between items. There are two typical criteria, including cosine measure and Pearson-correlation, to construct the similarity matrix $S$. Let $\mathbf{v_i}\in \mathbf{R}^m$ denote the interactions between the item $v_i$ and users. Then the consine similarity between two items $v_i$ and $v_j$ can be computed as 
\begin{equation}
\label{eq:knn:consine}
	S_{ij} = \frac{\mathbf{v}_i\mathbf{v}_j}{||\mathbf{v}_i||_2*||\mathbf{v}_j||_2}
\end{equation} 
For Pearson-correlation, the similarity is
\begin{equation}
\label{eq:knn:pearson}
	S_{ij} = \frac{\sum_{k\in \{1:m\}}|\mathbf{v}_{ik}-\bar{v}_i||\mathbf{v}_{jk}-\bar{v}_j|}{\sqrt{\sum_{k\in \{1:m\}}(\mathbf{v}_{ik}-\bar{v}_i)^2}\sqrt{\sum_{k\in \{1:m\}}(\mathbf{v}_{jk}-\bar{v}_j)^2}}
\end{equation}
where $\bar{v}_j$ is mean of all elements in $\mathbf{v}_j$.

After that, the similarities between the user $u_i$ and items can be modeled as 
\begin{equation}
F(u_i) = \mathbf{u}_iS
\end{equation}
Finally, the most similar items except those which have been interacted by users are kept as recommendations. Typically, to improve accuracy and reduce computational cost, only $K$ most similar items are kept for each item, that makes $S$ be a sparse matrix where each row contains $K$ non-zero entries.
In our work, we build the KNN of blogs based on the user-blog follow relations, and predict what blogs a user would follow based on the blogs that the user has followed.

\subsection{Factorization Machines}
Factorization machines (FM) have been demonstrated as a state-of-the-art generic framework for response prediction, such as personalized ranking~\cite{Rendle2009}, collaborative filtering~\cite{Rendle2012FML}, click prediction~\cite{Rendle2012SNC}, etc. It can mimic different kinds of recommender models by adjusting the input data. Formally, let $(\mathbf{x}\in\mathcal{R}^\ell,y)$ be an instance, where $\mathbf{x}$ is the feature vector, $y$ is the corresponding response, $f$ is the number of features. Suppose there are $\ell$ instances, and then we have a data set $(X,Y)=\{\mathbf{x}_i,y_i\}_{i=1}^{\ell}$. FM aims to build a polynomial function $F$, such that each feature vector can be mapped to the corresponding response, by minimizing some loss functions.
\begin{equation}
\min_{F}\ell(Y, X|F)
\end{equation}
Specifically, $d$-order polynomial function can be defined as
\begin{equation}
\label{eq:polynomial}
F(\mathbf{x})=w_0 + \sum_{i=1}^{n}w_ix_i + \sum_{g=2}^{d}\sum_{i_1=1}^{n}\cdots\sum_{i_g=i_{g-1}+1}^{n}\Big(\Pi_{j=1}^{g}x_{i_j}\Big)w_{\{i_{j}\}_{j=1}^{g}}
\end{equation}
where $w_0$ is the zero-order parameter, $w_i$ is the first-order parameter of the $i$-th feature and $w_{\{i_{j}\}_{j=1}^{g}}$ is the corresponding high-order parameters. Different from the traditional polynomial regression, FM factorizes the model parameters of high-order features into low-rank representations, where 
\begin{equation}
\label{eq:fm}
w_{\{i_{j}\}_{j=1}^{g}}=\sum_{g=1}^{d}\Pi_{j=1}^{g}\mathbf{z}_{i_j,f}^{(g)}
\end{equation}
and $\mathbf{v}$ is a latent feature vector with $k$ dimensions. In real-world applications, we typically set $d=2$. In this case, we can rewrite FM as
\begin{equation}
\label{eq:fm2}
F(\mathbf{x})=w_0+\sum_{i=1}^{n}w_ix_i+\sum_{i=1}^{n}\sum_{j=i+1}^{n}\mathbf{z}_i\mathbf{z}_j^Tx_ix_j
\end{equation}
As $F$ is convex with respect to $w_0$, $w_i$ and $\mathbf{v}_i$ individually, we can adopt gradient based methods to learn model parameters. From machine learning point of view, this factorization technique reduces the model complexity, where high-order parameters are represented in compact formats, and hence the risk of overfitting. From recommendation system point of view, since the similarity between different features can be captured by their dot-product scores, FM is also representative for latent factor models in collaborative filtering.  

One typical application of FM is the standard matrix factorization (MF)~\cite{Koren2009MFT}. In MF we typically factorize the matrix $R$ into two low-rank matrices $U$ and $V$, or  
\begin{equation}
\label{eq:mf}
R_{pq}=b_0+b_p+b_q+U_pV^T_q
\end{equation}
where $b_0$, $b_p$, $b_q$ are the global, user and item biases respectively and $U_p$ and $V_q$ are the corresponding user and item factors, respectively. Here what we observe are only users and items as well as users' responses on items. Thus, each instance in MF can be represented as a triple, which contains user and item indices as well as the response. Under the FM framework, we represent each instance as a sparse vector with $m+n$ dimensions. There are only two non-zero entries of which values are $1$, corresponding to the user and item indices respectively. Suppose the user index is $p$ and the item index is $q$, and then the prediction of the instance is represented as 
\begin{equation}
\label{eq:fm:mf}
f(\mathbf{x})=w_0+w_p+w_q+\mathbf{z}_p\mathbf{z}^T_q
\end{equation}
This equation is exactly the same as Eq.(\ref{eq:mf}).

\subsection{Modeling App Usages}
As shown above, these two models both rely on users' interactions on items. If users' interaction records are limited, we may hard to infer users' preference and fail to provide accurate predictions. However, if one user has related auxiliary information, such as the app usage logs as we shown in Section~\ref{sec:appblogdata}, we may use this information to improve our modeling on users' interests and thus boost the recommendation performance. We describe how to utilize the two algorithms to build recommendation models using app usage logs. The motivation is to embed the correlation between apps and blogs into the model building process. 

For KNN, similar with building the similarities between blog, we also build the similarity between apps and blogs $S_a\in \mathbf{R}^{c*n}$ using Eq.(\ref{eq:knn:consine}) or Eq.(\ref{eq:knn:pearson}) and then the combined prediction for the user $u_i$ is computed as
\begin{equation}
	\alpha A_iS_a + (1-\alpha) \mathbf{u}_iS
\end{equation}
where $\alpha$ is the trade-off parameter which can be learned from a hold-out dataset during the training process. After that, most similar blogs of the user are selected as recommendations.

For matrix factorization, the extension is also straight-forward. Besides modeling the interaction between users and items, we also model the correlations between users' used apps and the followed blogs. Formally, we turn the Eq.(\ref{eq:mf}) into
\begin{equation}
	R_{pq}=b_0+b_p+b_q+\sum_{A_{pi}\neq 0}b_i+U_pV^T_q+\sum_{A_{pi}\neq 0}W_iV^T_q
\end{equation}
where $W_i$ represents the latent factors of the $i$-th app and $b_i$ is the corresponding bias. We can also transform this equation using the FM formula. For each training instance, besides the user and item indices, we can also append apps used by users and make each training example as a vector with length $m+n+c$. The non-zero values in the vector represent the indices of the corresponding user, blog and all apps used by the user. Then we obtain
\begin{equation}
\label{eq:fm:mf}
f(\mathbf{x})=w_0+w_p+w_q+\sum_{A_{pi}\neq 0}w_i+\mathbf{z}_p\mathbf{z}^T_q+\sum_{A_{pi}\neq 0}\mathbf{z}_i\mathbf{z}^T_q
\end{equation} 
We observe that, the correlations between apps and blogs are built through $\sum_{A_{pi}\neq 0}\mathbf{z}_i\mathbf{z}^T_q$.

\section{Experiments}\label{sec:experiment}
In this section, we present a series of experiments that evaluate the performance of the proposed recommendation approaches for blog recommendation. We first give a detailed description of the datasets including both blog following and app usage, and the settings of our experiments. Then, we compare and discuss the performance of the proposed approaches against a set of baselines. The purpose of our experiments is to validate the effectiveness of exploiting apps for improving blog recommendation in Tumblr.

\subsection{Datasets}\label{sec:dataset}
As mentioned in Section~\ref{sec:appblogdata}, our blog following dataset is a sampled set of user-blog following bipartite in Tumblr. The dataset contains over 2 million users and over 20 million blogs with overall more than 5 billion follow events between them. The data sparseness is severely at 0.01\%, which illustrates the obvious challenge of blog recommendation. The app usage dataset is collected internally for the users who are involved in the Tumblr dataset. To reduce the noise in the dataset, we only keep the top 10 apps that the user used most frequently in the past week. In total, we have more than tens of thousands apps in the dataset. For business confidentiality, we cannot present more detailed information of the apps in our collection. Note that as described in Section~\ref{sec:model}, we take the apps that each user has as the user side features in FM. Therefore, in FM we learn not only latent factors of users and blogs, but also latent factors of apps.

\subsection{Experimental Setup}\label{sec:setup}
In the blog follow dataset, for each user we split his followed blogs into 80\% for training and 20\% for testing. In the test set, for each user we include a set of randomly selected blogs, which are in the number of five times as many as the ground truth blogs. The evaluation is based on how accurately we recommend the ground truth blogs against the randomly selected blogs for individual users. For example, if one user has one ground truth blog in the test set, we will include another five randomly selected blogs in the test set for this user. As a result, a random recommendation approach for this user would result in a probability of 1/6 for getting the ground truth blog on top.

To measure the recommendation performance, we adopt standard metrics, precision at top-N (P@N) and mean reciprocal rank (MRR)~\cite{Voorhees2000OOT}. P@N measures the ratio of relevant blogs we recommend in the top-N list, and MRR is the inverse of the rank of the first relevant blog, meaning how early we can make a relevant recommendation. In short, the larger P@N and MRR, the better the recommendation performance.

\textbf{Baselines.} In our experiments, we compare the performance of app-based FM to a few baseline approaches as listed below:
\begin{itemize}
\item \textbf{POP.} A naive and non-personalized approach that recommends the globally most popular blogs to every user. We generate the most popular blogs based on the number of follows of each blog in the training set.
\item \textbf{ItemCF.} This is the conventional item-item CF algorithm~\cite{Sarwar2001ICF} that recommends similar blogs to a user based on the blogs that the user has followed. In our experiment we tuned the neighborhood size to be 50, which gives nearly the best performance of this approach.
\item \textbf{MF.} This is the standard matrix factorization approach that factorize user-blog follow relations into latent factors, which are used further to predict the relevance between users and blogs. Note that MF can be seen as a simple version of FM, in which only user-blog relations are used.
\end{itemize}

\textit{Additional Settings.} For both MF and FM we set the number of latent factors to be 5, which is in consideration of both accuracy of the prediction and the workload of the computation, as it would largely increase both the space and time complexity if we were to employ larger latent space. In addition, for both MF and FM we need to sample negative examples in order to train the models. For this reason, in each case, we randomly selected the same number of blogs as the followed ones for each user to be the negative examples. In this work, we did not further tune the ratio between positive and negative examples, since the main focus is to validate the usefulness of mobile apps for blog recommendation, while not being to tweak the latent factor models.

\textit{Implementation.}
We implemented all the approaches in the company's internal Hadoop Map-Reduce cluster with more than 1000 machines, in order to scale up for the large amount computation. For POP and ItemCF, the implementation is straightforward, and only requires a few map-reduce jobs. Similar implementation can be found in literature~\cite{Schelter2012SSN}. For the latent factor models, the implementation with map-reduce is more sophisticated. We refer readers who have interest to our previous work~\cite{Zhong2015BDU} which elaborates the systematic workflows of the implementation.

\subsection{Performance}\label{sec:performance}
We show the overall performance of app-based FM and all the baselines in Table~\ref{tab:result}. First, we can see that the popularity-based blog recommendation is substantially outperformed by all the personalized approaches, indicating that personalization is indeed critical for the blog recommendation problem.
Second, we see that by exploiting app usage information, app-based FM achieves substantial improvement over ItemCF and MF, both of which only rely on the user-blog follow data. The improvement is over 2.3\% in P@1, 2.2\% in P@5 and 3.2\% in MRR, and all the results are statistically significant. This observation indicates that the app usage data from the mobile users do carry information that reflects the user's interest, and thus, can benefit for blog recommendation in mobile devices.

\begin{table}[t]
\caption{Performance comparison between app-FM and baseline approaches.}
\begin{center}
\begin{tabular}{|l|c|c|c|c|}
\hline
&P@1&P@5&MRR \\ 
\hline
POP&0.225&0.146&0.354 \\
ItemCF&0.806&0.614&0.865 \\
MF&0.748&0.542&0.839 \\
app-FM& \textbf{0.825} & \textbf{0.628} & \textbf{0.893} \\
\hline
\end{tabular}
\label{tab:result}
\end{center}
\end{table}

\begin{figure}[p]
\begin{small}
\centering 
\subfigure[P@1]{\scalebox{1.0}{\includegraphics[width=\columnwidth]{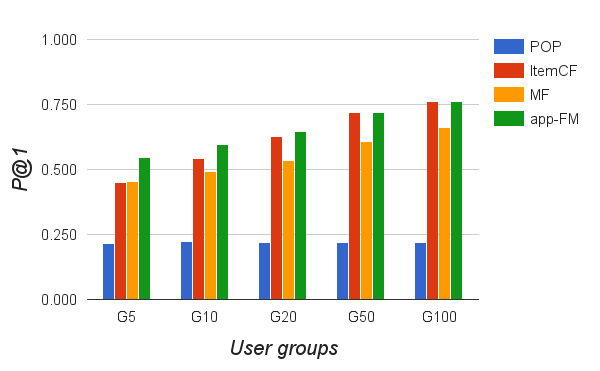}}\label{fig:user:p1}}
\subfigure[P@5]{\scalebox{1.0}{\includegraphics[width=\columnwidth]{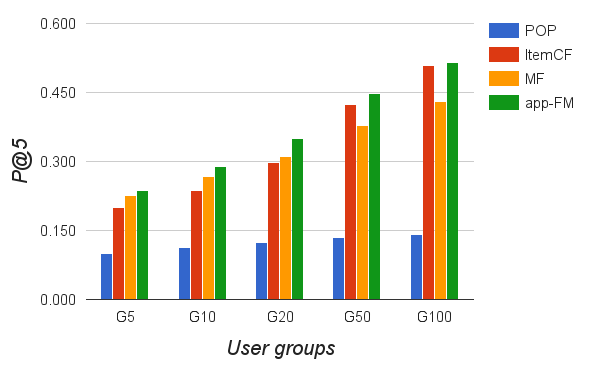}}\label{fig:user:p5}}
\subfigure[MRR]{\scalebox{1.0}{\includegraphics[width=\columnwidth]{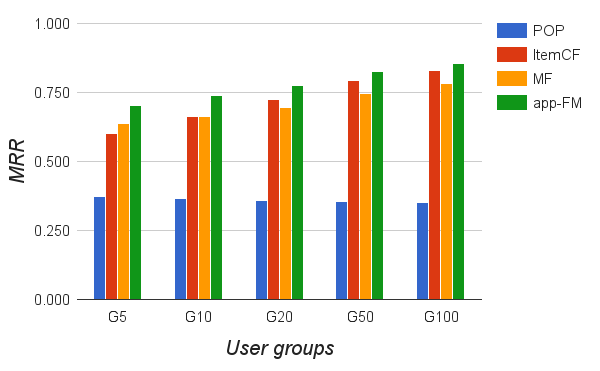}}\label{fig:user:mrr}}
\caption{Performance on users with different number of followed blogs.} \label{fig:usergroup}
\end{small}
\end{figure}

We further investigate the impact of the followed blogs per user on the performance of each recommendation approach. In this respect, we split the users in the test set based on the number of blogs that the user follows, and measure the performance for each user group. Specifically, we use ``G10" to denote the user group in which each user has less than 10 followed blogs in the training set. 
We show the comparison results for P@1, P@5 and MRR, respectively, in Figure~\ref{fig:usergroup}, in which we come into three observations.
First, the POP approach, as known to be non-personalized, is not strongly influenced by the number of user followed blogs. i.e., P@1 remains similar for different user groups, while P@5 and MRR are not consistent in showing the performance pattern.
Second, in contrast to POP, all the personalized approaches (both baselines and app-based FM) show clear impact from the number of user followed blogs. It is consistent that the performance of all these approaches decreases as the users follow less blogs, which is intuitive to our understanding that the recommendation performance would be degraded for cold start cases.
Third, we also notice that although the performance degrades when the number of user followed blogs decreases, app-based FM achieves relatively higher improvement for the users with less followed blogs. For example, the improvement of app-FM over ItemCF on the user group G100 is around 2.9\%, while it raises up to 16.6\% for the user group G5. This result again indicates that app-based FM is particularly effective for improving the cold start cases in blog recommendation.

\section{conclusion}\label{sec:conclusion}
We address the blog recommendation in mobile domain by leveraging app usage data into the Factorization Machine framework. Through our experiments on the Tumblr blog following dataset, we show that exploiting the app usage data in FM leads to a superior performance of blog recommendation compared to other baselines, especially for the cold start users who have very limited number of followed blogs. Our future work involves a few directions. First, we would like to deploy app-based FM approach into online test, and compare it with production baselines. Second, we are interesting in exploiting different types of user actions on apps to derive more discriminative user interests, which can be further used for improving blog recommendation. Third, as this work is the beginning of our contributions to improving recommendation application in mobile devices, we also like to explore novel application domains and novel features such as context and network structures.

{
\bibliographystyle{abbrv}
\bibliography{source}
}
\end{document}